# CALCULATION OF SPECTRA OF THE LATTICE AND SURFACE VIBRATIONS OF ORGANIC NANO-CRYSTALS

M.A.Korshunov[*]

*L.V. Kirensky Institute of physics, Siberian branch of the Russian Academy of Sciences, 660036 Krasnoyarsk, Russia*

**Abstract**. Calculations of frequencies of the lattice and surface oscillations of organic nano-crystals are carried out. As the sample the para-dichlorobenzol has been chosen. Change of spectra of oscillations from the sizes of nano-particles is found. It is shown that with the reduction of the sizes of nano-particles a spectrum of the surface oscillations prevailing. Calculations have shown that for the correct interpretation of the observational spectra it is necessary to consider orientation disorder of the surface molecules and presence of vacancies in nano-particle volume.

Organic crystals find the increasing application in the molecular electronics (at making of microcircuits, and in storage devices). Nano-technologies are thus used. Therefore examination of organic molecular nano-crystals represents practical interest. As it is scored in a number of operations [1] physical properties of nano-particles constructed of atoms differ from properties of crystals of the major sizes. Change of physical properties because of change of the sizes of crystal grains for the structures constructed of organic molecules, apparently, should be observed at the major sizes, than for the structures constructed of atoms. In operation [2] para-dichlorobenzol spectra are gained Raman at change of the sizes of crystals from 5μ and less 1μ (~500nm). Shift of lines towards excited and their broadening is observed. Appearance and increase of intensity of additional lines. For an explanation of observable changes of frequencies of lines calculations of spectra of the lattice oscillations on a method the Dyne [3] have been carried out. Minimisation on energy was done at a various arrangement of molecules and their quantity. Change of parametres of a lattice was thus considered at change of the sizes of nano-crystals [4]. Change of a relation volume/surface was considered at reduction of the sizes of nano-particles.

In present work, the calculations of the lattice oscillations of a nano-particle of para-dichlorobenzol free of defects and calculations of the superficial oscillations are carried out. In Figs. 1-3 histograms of frequency spectra are shown at change of the sizes of nano-crystals.

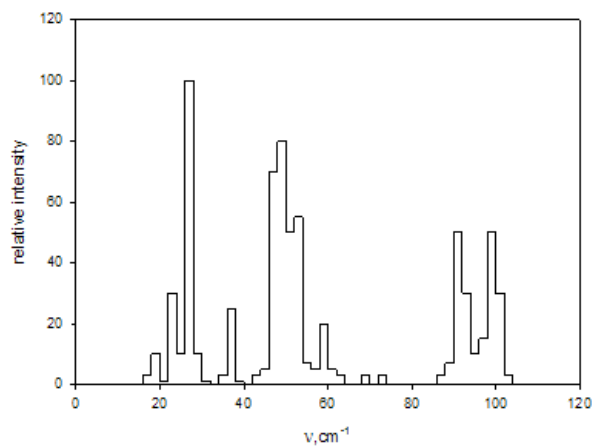

**Fig. 1.** The histogram of a spectrum of a nano-particle of para-dichlorobenzol (ratio volume/surface >> 1).

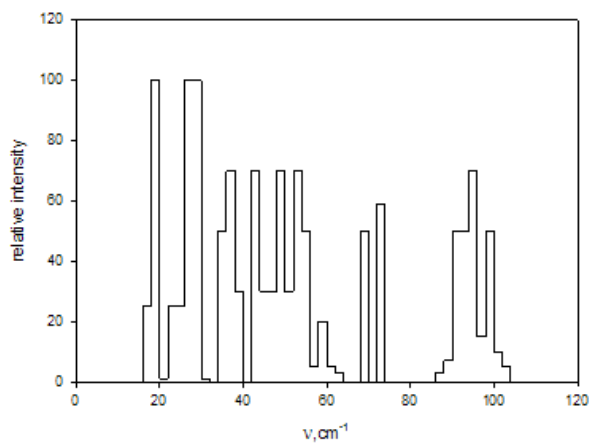

**Fig. 2.** The histogram of a spectrum of a nano-particle of para-dichlorobenzol (ratio volume/poverhnost = 1).

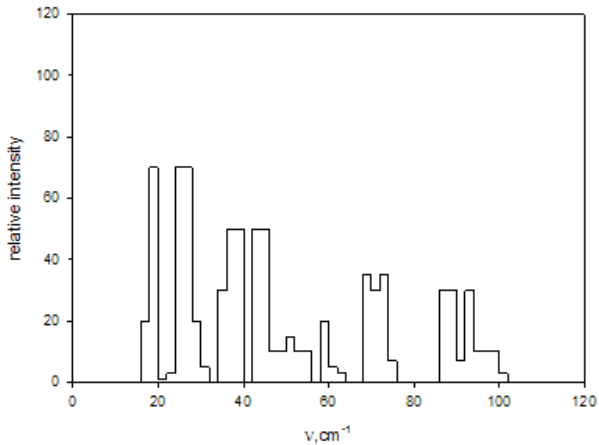

Fig. 3. The histogram of a spectrum of a nano-particle of para-dichlorobenzol (ratio volume/surface << 1).

In the first Figure a relation volume/surface>>1, in the second Figure volume/poverhnost=1 and on the third volume/surface<<1. As we see, there is a spectrum change, as on the relative intensity of lines of a spectrum, and values of frequencies. To reduction of the sizes of crystals there is an increase of a role of a spectrum of the superficial oscillations in the net spectrum of a nano-particle. In field from 20 cm$^{-1}$ to 100 cm$^{-1}$ the change in a spectrum corresponds to changes in the observational spectrum at change of the sizes of particles. But during too time occurrence of lines of a spectrum below 15 cm$^{-1}$ and lines in the field of 80 cm$^{-1}$ is not observed. That is the spectrum of nano-particles free of defects not completely corresponds to a spectrum observed in experiment (Fig. 4). Therefore, calculation has been carried out at orientation disorder of the superficial molecules. It has given occurrence of additional lines in field below 15 cm$^{-1}$ and to a broadening of other lines of a spectrum, but in the field of 80 cm$^{-1}$ of lines has not appeared. At the account of vacancies of lines in this field have appeared. In Fig. 4 the observational spectrum of a nano-particle of para-dichlorobenzol (~500nm) and the calculated histogram of a spectrum is shown at the account of orientation disorder of the superficial molecules and presence of vacancies. Apparently, the settlement spectrum is close to the observational spectrum.

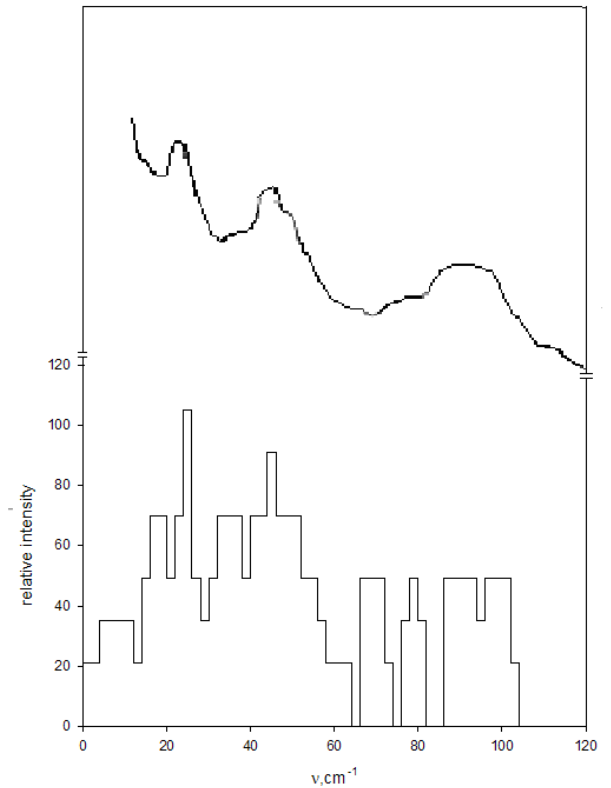

Fig. 4. The observational spectrum of a nano-particle of para-dichlorobenzol (~500nm) and the calculated histogram of a spectrum at the account of orientation disorder of the superficial molecules and presence of vacancies (ratio volume/surface << 1).

Thus, it is shown, that with reduction of the sizes of nano-particles the spectrum of the superficial oscillations becomes prevailing. For the correct interpretation of the observational spectra, it is necessary to consider orientation disorder of the superficial molecules and presence of vacancies in nano-particle volume.

### References

[1]. A.I.Gusev. Usp.Fiz.Nauk 168, 1, 55 (1998).
[2]. arXiv:physics/0610245 v1, 26 Oct 2006.
[3]. P.Dean, J.L.Martin, Proc. Roy. Soc. 259,409 (1960).
[4]. arXiv:cond-mat/0709.0066 2007